# Parallel and Real-Time Post-Processing for Quantum Random Number Generators

Xiaomin Guo[1,2], Mingchuan Wu[1], Jiangjiang Zhang[1], Ziqing Wang[1], Yu Wang[2,*] and Yanqiang Guo[1,3,*]

[1]Key Laboratory of Advanced Transducers and Intelligent Control System, Ministry of Education, College of Physics and Optoelectronics, Taiyuan University of Technology, Taiyuan 030024, China
[2]State Key Laboratory of Cryptology, Beijing 100878, China
[3]State Key Laboratory of Quantum Optics and Quantum Optics Devices, Shanxi University, Taiyuan 030006, China

*Corresponding author: Yanqiang Guo (guoyanqiang@tyut.edu.com); Yu Wang (wangy@sklc.org)

This work was supported in part by the National Natural Science Foundation of China under Grant 61875147, Grant 61671316, and Grant 61731014, in part by the Key Research and Development (R&D) Program of Shanxi Province (International Cooperation) under Grant 201903D421049, in part by the National Key Research and Development Program of China under Grant 2016YFA0302600 and Grant 2018YFA0306404, in part by the Natural Science Foundation of Shanxi Province under Grant 201701D221116 and Grant 201801D221182, in part by the Scientific and Technological Innovation Programs of Higher Education Institutions in Shanxi under Grant 201802053 and Grant 2019L0131, in part by the Program of State Key Laboratory of Quantum Optics and Quantum Optics Devices under Grant KF201905, in part by the Research Project Supported by Shanxi Scholarship Council of China under Grant 2017-040, and in part by the Shanxi 1331 Project Key Innovative Research Team.

**Abstract** Quantum random number generators (QRNG) based on continuous variable (CV) quantum fluctuations offer great potential for their advantages in measurement bandwidth, stability and integrability. More importantly, it provides an efficient and extensible path for significant promotion of QRNG generation rate. During this process, real-time randomness extraction using information theoretically secure randomness extractors is vital, because it plays critical role in the limit of throughput rate and implementation cost of QRNGs. In this work, we investigate parallel and real-time realization of several Toeplitz-hashing extractors within one field-programmable gate array (FPGA) for parallel QRNG. Elaborate layout of Toeplitz matrixes and efficient utilization of hardware computing resource in the FPGA are emphatically studied. Logic source occupation for different scale and quantity of Toeplitz matrices is analyzed and two-layer parallel pipeline algorithm is delicately designed to fully exploit the parallel algorithm advantage and hardware source of the FPGA. This work finally achieves a real-time post-processing rate of QRNG above 8 Gbps. Matching up with integrated circuit for parallel extraction of multiple quantum sideband modes of vacuum state, our demonstration shows an important step towards chip-based parallel QRNG, which could effectively improve the practicality of CV QRNGs, including device trusted, device-independent, and semi-device-independent schemes.

## 1. Introduction

For applications where security and unpredictability are of utmost importance, true random number generators (TRNGs) play a heavy role compared to its pseudo-random counterparts. Quantum random number generators (QRNGs) exploits inherent uncertainty essence of quantum physics to generate unpredictable, non-reproducible true random numbers that can be applied to high-security-requirements areas, including quantum key distribution, secure communication, etc.

Among various QRNGs, schemes based on CV quantum fluctuations, including device trusted, device-independent and semi-device-independent schemes [1-5], have become promising for its convenience of state preparation, insensitivity of detection efficiency, high measurements bandwidth, and stable and compact optical setup. In such schemes, the model between the ideal random number exploiting process and defects in practical implementation can be established for information-theoretical provable security. However, generic to most TRNGs, who can be classified as serial type, potential limitations exist in these systems. For example, such serial structure of system always faces the rate bottleneck, especially in the step of post-processing. That is, as for high throughput of true random numbers, system has stringently demands on electronic components and circuits and eventually leads to expensive cost.

Parallel random bit generation methods could dramatically enhance the generation rate and scalability of RNG by producing random bits simultaneously from multiple paths. Parallel pseudorandom number generation algorithms have been developed [6, 7] and well applied in the field of mathematical modelling. Parallel hardware pseudorandom number generation methods based on cellular automata have been demonstrated [8-10]. Physical random number generators based on superluminescent LED [11, 12] and chaotic laser have also been reported [13, 14]. In our previous work, we have proposed parallel QRNG based on quadrature measurement of vacuum fluctuations [15]. Optical and electronic elements involved are employed cost-effectively and integratable based upon the existing technology [16-18]. The last but the most important element, the randomness extraction, is realized based on the security information theoretical provable Toeplitz hash extractors.

Real-time Toeplitz hash post-processing based on field-programmable gate array (FPGA) has been reported recent years [19-23]. Compared to offline post-processing [24, 25], FPGA-based randomness extractor (RE) extracts and outputs true random numbers while ADC acquiring raw random bits sequences, which is desiderated in practical cryptographic applications. This type of work has achieved very high random number generation speed. Real-time productive rate up to 3.2 Gbps [19], 6 Gbps [20, 21] and 2.9 Gbps [22] have been reported recently. Note that the performance of their high productive rate mainly depends on high ADC acquisition rate and FPGA calculation speed, resting with high economic cost. ADC acquisition rate reaches up to 1 GSps in some works [19-22] and fully exploitation of hardware sources have not been concerned in these works. In fact, the critical component for a real-time QRNG, the post-processer is the focal point of the implementation cost since the price of ADC rise sharply over with sample rate and FPGA with quantity of look-up tables (LUT). In this way, elaborate layout and efficient utilization of hardware computing resource in the realization of RE are crucial important.

In this work, we experimentally investigate real-time parallel post-processing of vacuum state-based QRNG and the scheme is shown in Fig. 1. In the respect of entropy source, multiple independent quantum sideband modes within bandwidth of one homodyne detection system are extracted simultaneously. By delicate designation of two-layer parallel pipeline algorithm and full exploitation of the parallel algorithm advantage of FPGA, common configuration ADC and FPGA successfully support the post-processing of the multichannel parallel QRNG and 8.24 Gbps real-time post-processing of tri-channel quantum random number extraction is realized. Logic source occupation for different scale and quantity of Toeplitz matrices is analyzed and finally 3-channel post-processing takes up about 61% of the logic resources in the FPGA, and other modules such as PCI-E interface control occupy about 30%. To test the validation and independence of each Toeplitz-hash extractor, the autocorrelation function (ACF) and cross-correlation function (CCF) of the random numbers before and after post-processing are compared firstly. Then NIST test on output of each channel is implemented independently. Furthery, the NIST test is used to check their combination output. All above tests are passed successfully and the effect of our parallel real-time Toeplitz-hash extractors is ensured. In order to ensure the robustness of the system in long time, we made NIST tests on sampled combination output random numbers evenly distributed within 10 hours and recorded the minimum pass rate. Kullback-Leibler divergence of these random numbers within the same process is analyzed.

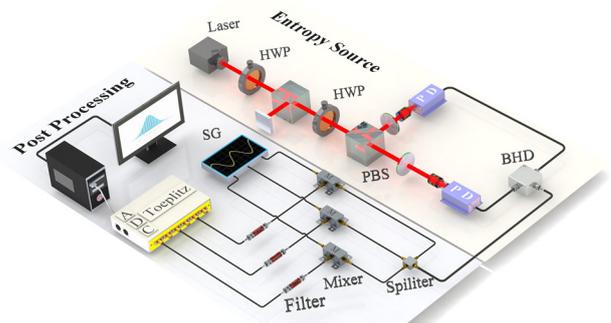

**Figure 1.** Experimental scheme for real-time parallel post-processing of vacuum state-based QRNG. The entropy source part is a use of vacuum-based scheme to generate a randomly fluctuating voltage signal. The post-processing part utilizes mixing, filtering, analog-to-digital conversion, and Toeplitz post-processing to obtain random bits. HWP, half wavelength plate; PBS, polarization beam splitter; BHD, balanced homodyne detector; SG, signal generator; ADC, analog to digital converters.

This represents an important step towards a chip-based parallel implementation of information theoretically secure RE, which could dramatically improve the generation rate and scalability of information theoretically secure QRNGs including device trusted, device-independent and semi-device-independent schemes [1-5].

## 2. Implementation

The overall structure of the tri-channel Toeplitz-hash extractors constructed in one FPGA is sketched in Fig. 2 (a), and the operation process of one of the three channels in inner layer is shown in Fig. 2 (b).

We present this work as following three parts:

### 2.1 Two-layer parallel structure

Compared with processors with serial computing architectures, FPGAs have the main advantage of parallel logical processing and integer calculations. In our work, two-layer parallel structure is constructed to realize the Toeplitz-hash post-processing for three sets of raw data. Strength of parallel algorithm in FPGA is fully utilized.

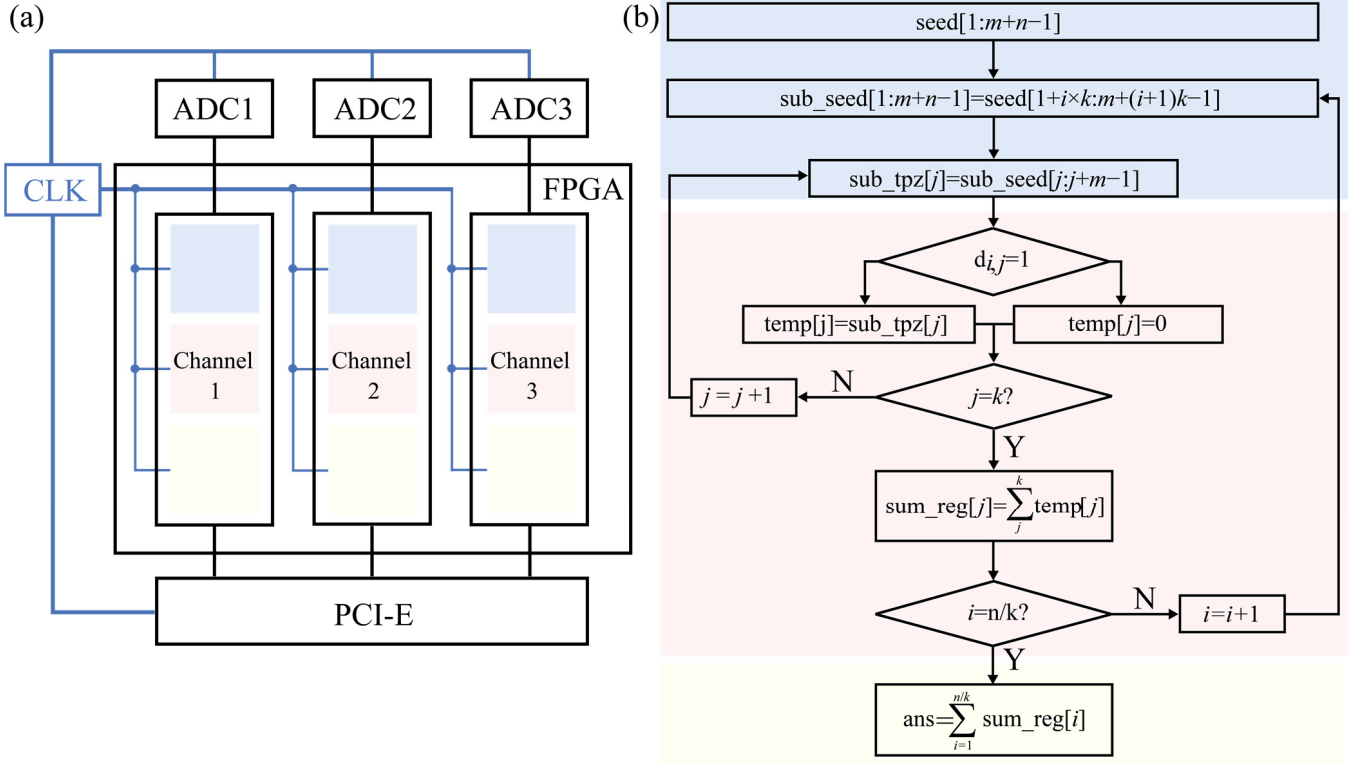

**Figure 2.** Part (a) shows the overall structure of the multi-channel process, and part (b) shows the operation process of one of the three channels. The three modules of the pipeline processing including matrix construction module, sub-matrix multiplication module and vector accumulation module are shown in the figure. Note that only the logic flow of Toeplitz post-processing is shown in the figure. The modules in the FPGA operate parallelly.

At the outer layer, extractors for the three sub-entropy sources are constructed independently and operate concurrently. The conditional minimum entropy is different for the three modes, due to the nonconstant gain spanning the balanced homodyne detector (BHD) bandwidth, so different Toeplitz matrix scales for the three channels are set. ADC acquisition of the three channels and real-time parallel processes in the FPGA are controlled by the same clock. The random numbers of the three channels are alternately mixed and transmitted to the PC in real time.

At the inner layer, the process of the Toeplitz matrix operation mainly consists of AND operation and XOR operation, which can be realized by the basic logic unit in the FPGA. For each quantum sideband mode, Toeplitz post-processing is implemented in a pipeline algorithm that consists of three modules, including a matrix construction module, a sub-matrix multiplication module and a vector accumulation module. Since the matrix sizes of the three Toeplitz-hashing extractors are different, in the following description of the specific algorithm, $m$, $n$ and $k$ are used to illustrate the general case.

At each a clock cycle of the FPGA, the multiplication between a sub-matric and a subsequence in each RE is calculated simultaneously. As shown in Eq. (1), the m×n Toeplitz matrix is divided into n/k (n is an integer multiple of k) sub-matrices with a size of m×k, so that the entire Toeplitz matrix processing is split into n/k steps. Each step is a multiplication calculation of m×k sub-matrices and k-bits raw random number sequence, which occupies one clock cycle. For each RE of a sub-entropy source, in one clock cycle, the FPGA completes the generation of a Toeplitz submatrix and the processing to the k-bit raw random number using the submatrix, and the ADC completes a sample collection. After n/k clocks, all n/k sub-matrices are used. The calculated m-bit vector is the result of post-processing of the entire Toeplitz matrix. During the n/k clocks, the ADC generates n-bits raw random numbers, and the FPGA processes the n-bit raw random number in real-time and output m true random numbers.

$$\begin{bmatrix} s_m & s_{m+1} & \cdots & s_{m+n-1} \\ s_{m-1} & s_m & \cdots & s_{m+n-2} \\ \vdots & \vdots & \ddots & \vdots \\ s_2 & s_3 & \cdots & s_{n+1} \\ s_1 & s_2 & \cdots & s_n \end{bmatrix} \times \begin{bmatrix} d_1 \\ d_2 \\ \vdots \\ d_{n-1} \\ d_n \end{bmatrix}$$

$$= \begin{bmatrix} s_m & s_{m+1} & \cdots & s_{m+k-1} \\ s_{m-1} & s_m & \cdots & s_{m+k-2} \\ \vdots & \vdots & \ddots & \vdots \\ s_2 & s_3 & \cdots & s_{k+1} \\ s_1 & s_2 & \cdots & s_k \end{bmatrix} \times \begin{bmatrix} d_1 \\ d_2 \\ \vdots \\ d_{k-1} \\ d_k \end{bmatrix} + \begin{bmatrix} s_{m+k} & s_{k+m+1} & \cdots & s_{m+2k-1} \\ s_{m+k-1} & s_{k+m} & \cdots & s_{m+2k-2} \\ \vdots & \vdots & \ddots & \vdots \\ s_{k+2} & s_{k+3} & \cdots & s_{2k+1} \\ s_{k+1} & s_{k+2} & \cdots & s_{2k} \end{bmatrix} \times \begin{bmatrix} d_{k+1} \\ d_{k+2} \\ \vdots \\ d_{2k-1} \\ d_{2k} \end{bmatrix}$$

$$+ \cdots + \begin{bmatrix} s_{m+n-k} & s_{m+n-k+1} & \cdots & s_{m+n-1} \\ s_{m+n-k-1} & s_{m+n-k} & \cdots & s_{m+n-2} \\ \vdots & \vdots & \ddots & \vdots \\ s_{n-k+2} & s_{n-k+3} & \cdots & s_{n+1} \\ s_{n-k+1} & s_{n-k+2} & \cdots & s_n \end{bmatrix} \times \begin{bmatrix} d_{n-k+1} \\ d_{n-k+2} \\ \vdots \\ d_{n-1} \\ d_n \end{bmatrix} = \begin{bmatrix} a_1 \\ a_2 \\ \vdots \\ a_{m-1} \\ a_m \end{bmatrix} \quad (1)$$

As shown in Fig. 2(b), the first stage of the pipeline algorithm for each Toeplitz-hashing extractor is the Toeplitz sub-matrix generation module, which is used to construct a Toeplitz sub-matrix per clock cycle. We first store the seed containing $m+n$-1 bits and make it change every clock cycle with a shift feedback register. Every clock cycle we select $m+k$-1 bits from the $m+n$-1 bit seeds as the elements of a submatrix. The method of selection is to choose from the ($i$-1)×$k$+1 bit to the $m+i$-1 bit in the total $m+n$-1 bit seeds at the $i$th clock cycle. The value of $i$ ranges from 1 to $n/k$. When constructing a submatrix, from the $j$th to the ($j+m$-1) th bit of the selected seeds fill the $j$th column of the submatrix. The value of $j$ ranges from 1 to $k$.

The second stage is a sub-matrix operation module. When realizing the calculation of a single submatrix, since there is no statement or function directly used in the binary matrix calculation in the hardware description language of FPGAs, the process of this operation needs to be disassembled. The method of the execution is as follow:

$$\begin{bmatrix} s_m & s_{m+1} & \cdots & s_{m+k-1} \\ s_{m-1} & s_m & \cdots & s_{m+k-2} \\ \vdots & \vdots & \ddots & \vdots \\ s_2 & s_3 & \cdots & s_{k+1} \\ s_1 & s_2 & \cdots & s_k \end{bmatrix} \times \begin{bmatrix} d_1 \\ d_2 \\ \vdots \\ d_{k-1} \\ d_k \end{bmatrix}$$
$$= \begin{bmatrix} s_m \times d_1 \\ s_{m-1} \times d_1 \\ \vdots \\ s_2 \times d_1 \\ s_1 \times d_1 \end{bmatrix} + \begin{bmatrix} s_{m+1} \times d_2 \\ s_m \times d_2 \\ \vdots \\ s_3 \times d_2 \\ s_2 \times d_2 \end{bmatrix} + \cdots + \begin{bmatrix} s_{m+k-1} \times d_k \\ s_{m+k-2} \times d_k \\ \vdots \\ s_{k+1} \times d_k \\ s_k \times d_k \end{bmatrix} \quad (2)$$

The multiplication operation in the brackets at the right side of the Eq. (2) is performed parallelly in the FPGA by AND gates, and the obtained column vectors are executed bitwise XOR by cascaded XOR gates to obtain the result of a single Toeplitz submatrix operation. When using Verilog to describe this stage, a generate-for block (Verilog language) is used to calculate each column vector on the right side of the Eq. (2). By judging whether a bit in the sample is 0 or 1, it is determined whether a column of 0 or a corresponding column of the sub-matrix is assigned to the two-dimensional array "temp". By using some cascaded XOR gates to bitwise XOR the obtained $k$ column vectors, we get the result of the single submatrix processing.

The third stage is register XOR module. In one clock cycle, when getting the result of a single submatrix processing, we store it in a register "sum_reg". After $n/k$ clocks, all $n/k$ sub-matrices are used, and the stored $n/k$ $m$-bit vectors are performed bitwise XOR. The $m$-bit vector is the result of post-processing of the entire Toeplitz matrix.

### 2.2 Resource allocation for the parallel algorithms

Being able to provide all the matrix elements is a prerequisite for performing matrix operations. Fortunately, compared to the general matrix consisting of $m×n$ matrix elements, the number of matrix elements contained in a Toeplitz matrix is only $m+n$-1. They can be stored inside the FPGA, without the need of external memory or communication between the FPGA and the external memory. According to the theoretical analysis of post-processing security by information theory, the dimensions of Toeplitz matrix, $m$ and $n$, should have a magnitude of $10^2$ at least [26]. The FPGA xc7k325t provides 407,600 slice registers, which fully meets the storage requirements for our parallel Toeplitz post-processing of three channels.

However, for large-scale matrix operations, the shortage of FPGA logic resources is often the biggest problem encountered in real-time implementation of Toeplitz-hashing extractor. Operations in FPGAs are typically implemented through look-up tables (LUT). The operation of a large matrix consumes a large number of LUTs. The FPGA chip xc7k325t can provide about 203,800 LUTs. Other modules such as ADC control and PCI-E interface need to use about 28.4% of all LUTs. The remaining resources may not be enough to build a Toeplitz matrix with a magnitude of $10^2$ directly.

The specific scales of the three Toeplitz matrixes, it is, the extraction ratio of quantum random number from each set of raw data needs to be determined. According to the information theory, extraction ratio of true random number in post-processing is constrained by the Leftover Hash Lemma [27]:

$$l < H_{min} - 2\log\frac{1}{\epsilon}, \quad (3)$$

where $H_{min}$ is the minimum entropy and $\epsilon$ is the hash security parameter which refers to the proximity of the distribution of extracted random number to the IID (independent and identical distribution). The ratio of number of elements in matrix rows to that in columns is determined by the minimum entropy together with the security parameter. While the upper limit of matrix dimension in an actual implementation is determined by the number of FPGA logic resources. Using a representative sample of 10 Mbits sourced from per sub-entropy source, and conditioned on best sampling range, the minimum entropy of the raw random numbers of the three channels are figured out as 12.9, 13.5 and 14.2 [15, 24, 28]. Based on above consideration and choosing a security parameter of $2^{-50}$, the matrix sizes of the three channels are set as 519×768, 548×768 and 581×768.

**Table 1.** LUTs Usage Quantity

| Module | LUTs Usage |
|---|---|
| ADC | 1083 |
| DDR3 | 12405 |
| PCI-E | 15814 |
| Toeplitz Extractor | 118001 |
| Trigger mode control | 28481 |

This table shows the LUTs usage ratio of main modules. The DDR3 control and trigger control module is not used in this work.

In order to reduce the use of resources furthermore, we decompose the calculation between a whole Toeplitz matrix and the raw random number sequence into calculations between several Toeplitz sub-matrices and shorter original random number subsequences at the cost of increasing the number of clock cycles required for the entire post-processing. On the other hand, for real-time QRNG, the random extractor should post-process immediately the digital signal sampled by the ADC.

In this way, the scale of the Toeplitz submatrix actually determines the logical resource occupation of the algorithm implemented in the FPGA. Backtrack to Eq. (1), the larger the value of $k$, the more logical resource is needed for each step. Whereas the value of $k$ cannot be set too small because of the real-time requirement — for each channel, within a certain time period (several clock cycles), the number of post-processing bits in the FPGA must be equal to the number of bits read by the ADC. In this work, the clock frequencies of ADC sampling and FPGA post-processing are both 240 MHz. The value of $k$ is set to 16, equaling to the resolution of the ADCs. Based on our pipeline algorithm designed for parallel multiple paths, the real-time post-processing is guaranteed.

Then the number of rows of each sub-matrix needs to be determined. In order for Toeplitz post-processing to have smaller hash security parameter, the number of rows in the submatrix should be as large as possible. In other words, the three Toeplitz calculation modules should occupy as much logical resources as possible. The Integrated Software Environment (ISE) is applied to complete the above design in the FPGA of model xc7k325t. Logical resource occupation with different numbers of channels is compared. It can be found that the LUTs occupied by PCI-E and other functional modules remain unchanged at approximately 28.4% of all FPGA LUTs as the number of parallel channels increases. Fig. 3 shows the occupancy of each module under the circumstance with different submatrix scales and different counts of channels. For three-way parallel post-processing and matrix scale of m=576, the Toeplitz matrix post-processing algorithm occupies approximately 61%. Based on above discussions, our FPGA can support three-way parallel computing with submatrix size of about 576× 16 synchronously and in real time. In this way, logic resource in a single FPGA is fully utilized in supporting Toeplitz matrix manipulation with better efficiency.

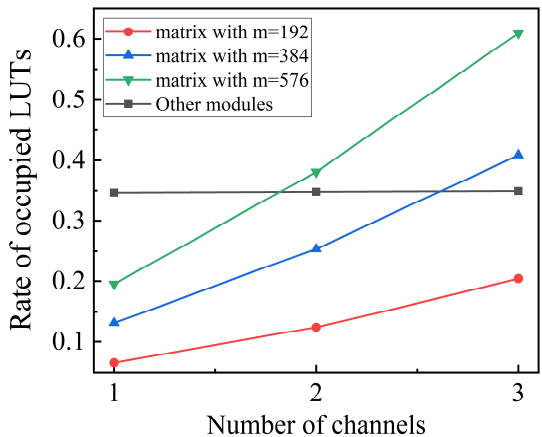

**Figure 3.** Logic resources consumption of the algorithm. The test is performed on the Xilinx 7k325t-fbg676, which has a total of 203,800 LUTs. The figure shows the percentage of LUTs used by Toeplitz operation module and other modules in the FPGA after mapping and routing when different channel counts and matrix scales are established.

### 2.3 Optimization of timing in FPGA

To sum up, facing limited logical resource in the FPGA, two-layer parallel architecture of three Toeplitz-hashing REs is designed to allocate the resource efficiently. Suitable small scales of Toeplitz matrix are set to ensure implementation of real-time post-processing of quantum random number at high speed.

Even so, when such an algorithm is implemented in FPGA, the fanout of signal lines induced by data sharing among multiple computing units may be large, which will affect the signals' timing, specifically, it will cause signal distortion, and finally affect data integrity. In this case, strong fanout drive capability is required for the logic resources in the FPGA and we need to find ways to reduce the number of fanouts to improve signal quality. In our work, a buffer for each module is invoked in the outer layer to mitigate the problems.

When the Toeplitz algorithm is implemented in the FPGA, we apply timing score to evaluate the quality of timing. Timing score is an indicator used to evaluate timing in ISE's report and characterizes the degree of completion of timing constraints. If the timing constraint is fully implemented, the timing score is 0. After using ISE's Smart Xplorer tool to map and route the above design, we find that the timing score value reported by ISE raises with the increase of matrix size and number of parallel channels, as shown in Table 2. From the timing report given by ISE, it can be seen that for a 576×16 submatrix, the fanout of the clock line is as high as $7.6 \times 10^6$, which we think is the cause of the high timing score. This situation may cause distortion of the clock signal and a bad data integrity.

**Table 2.** Timing Score for Different Matric Scales

| Submatrix Scale | Timing Score Before Optimization | Timing Score After Optimization |
| --- | --- | --- |
| 192×16 | 767 | 715 |
| 384×16 | $4.3 \times 10^4$ | $1.1 \times 10^4$ |
| 576×16 | $7.6 \times 10^6$ | $2.8 \times 10^4$ |

We tested the effect of adding buffers at different submatrix scales. The first column of the table is the scale of the submatrix. The second column is the timing score before adding buffers and the last column is the timing score after adding buffers.

In our work, global buffers of the FPGA are added to reduce clock fanout and thus improve the timing quality. Buffer is one of the internal clock resources of the FPGA. By default, each FPGA's input signal is connected to an Input Buffer as a driver. Here global buffers are added. Their input can be a clock generated by a phase-locked loop, an output of other buffers, an internal general signal line, etc. Global buffer's output can be connected to each clock input point on the chip. Xilinx 7 series FPGAs have 32 global buffers. We set them as single-ended input mode and connect their input ports to the output port of the clock's Input buffer. Then we use the Global Buffers to control the ADCs, each Toeplitz post-processing module and other functional modules. After the buffer is increased, the fanout allocated to each buffer will be much smaller than the original fanout, thereby reducing the requirements for the buffer's driving capability.

As a result of the improvement, the clock fanout of each Toeplitz post-processing module is approximately 8000, and the timing score is reduced from $7.6\times10^6$ to $2.8\times10^4$. Furthermore, we verified the impact of the reduction in timing score on signal integrity. Compared with the same post-processing process established in MATLAB, the bit error rates before and after increasing the buffer are $1.02\times10^{-8}$ and $3.73\times10^{-9}$, respectively.

## 3. Testing and analysis

Fig. 4 is the frequency spectrum diagram of the output voltage of the homodyne photodetector. Corresponding to the three-channel parallel post-processing of the FPGA, we extract three quantum sideband frequency modes with a bandwidth of 120 MHz within the measurement bandwidth. The center frequencies of the selected sidebands are 200 MHz, 600 MHz and 1 GHz respectively. The SNR of the noise fluctuations account for a large proportion of the three bands is greater than 10 dB, to ensure that quantum measurement results relative to the effects of classical noise.

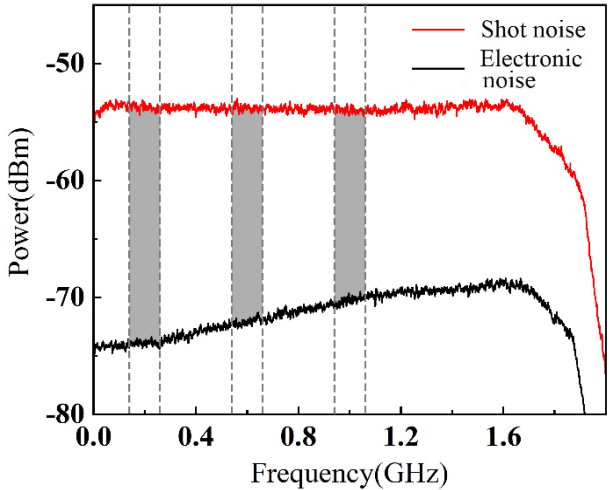

**Figure 4.** The black line is the power spectrum of the electronic noise, the red line is the power spectrum of the shot noise, and the gray bands are quantum entropy source selected by the mixer and the low pass filter.

To ensure the independence between the channels, autocorrelation and cross-correlation before and after post-processing of the random numbers are tested. Every 16-bit binary number is converted to a decimal number. As shown in Fig. 5, after post-processing, each ACF (autocorrelation function) has the shape of an impulse function. In areas where time is not zero, the ACF fluctuation range of the three channels reduces from about ±0.02 to ±0.002. The cross-correlation function (CCF) fluctuation range of the three channels reduces from about $\pm1.7\times10^{-3}$ to $\pm3\times10^{-4}$. This shows that random bits can be generated with high quality in each frequency band, and the correlation between the three channels can be ignored. In order to verify the randomness of the random bits strictly, we firstly collected the output sequences of three channels separately and performed 15 statistical tests using the standard NIST Statistical Test Suite. The significance level is set as α = 0.01 and 1000 sequences with 1 Mbits are subjected to the test. Then cumulated sequences from mixing of the three outputs from the three paths are tested under the same conditions. As shown in Fig. 6, all P values from the four tests are greater than 0.01, and the minimum pass rate for each statistical test is also within the confidence interval of 0.9803 to 0.9942. These tests identify the reliability of each Toeplitz-hashing extractor for every quantum sideband mode and demonstrate the randomness of the ultimate random numbers generated by the parallel QRNG.

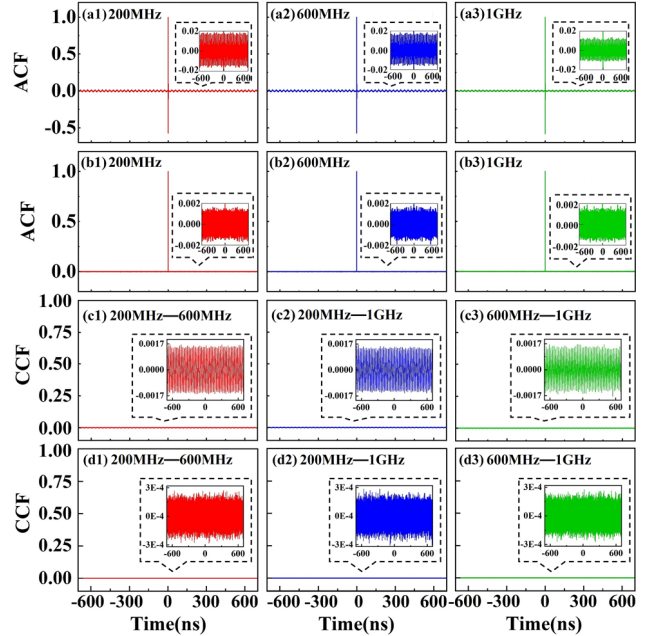

**Figure 5.** (a1)-(a3) is the autocorrelation of the raw random numbers from the three channels, (b1)-(b3) is the autocorrelation of the random numbers after post-processing; (c1)-(c3) is the cross-correlation of the raw random numbers of every two channels, (d1)-(d3) is the cross-correlation of the post-processed random numbers of every two channels.

In addition, in order to ensure the robustness and security of the system for future practical applications, operation stability tests on the QRNG are performed. Two tests are used to measure the randomness of the final quantum random numbers in long-term operation. The first one is NIST tests [29] to long-term sampling [30]. We kept the QRNG working continuously for more than 10 hours and ran a timing sampling program on the PC host computer. After the first collection, random numbers generated in real-time are collected every 450s (8 times per hour) with a random sequence of 1G bits gathered each time, and we kept sampling for a total of 10 hours. Then standard NIST Statistical Test Suite is used to all 80 samples collected within the 10 hours, and the minimum test pass ratio of each group is recorded. Long term acquisition of NIST test results is shown in Fig. 7, which shows average values of the minimum test pass ratio of the eight sets of random numbers collected per hour.

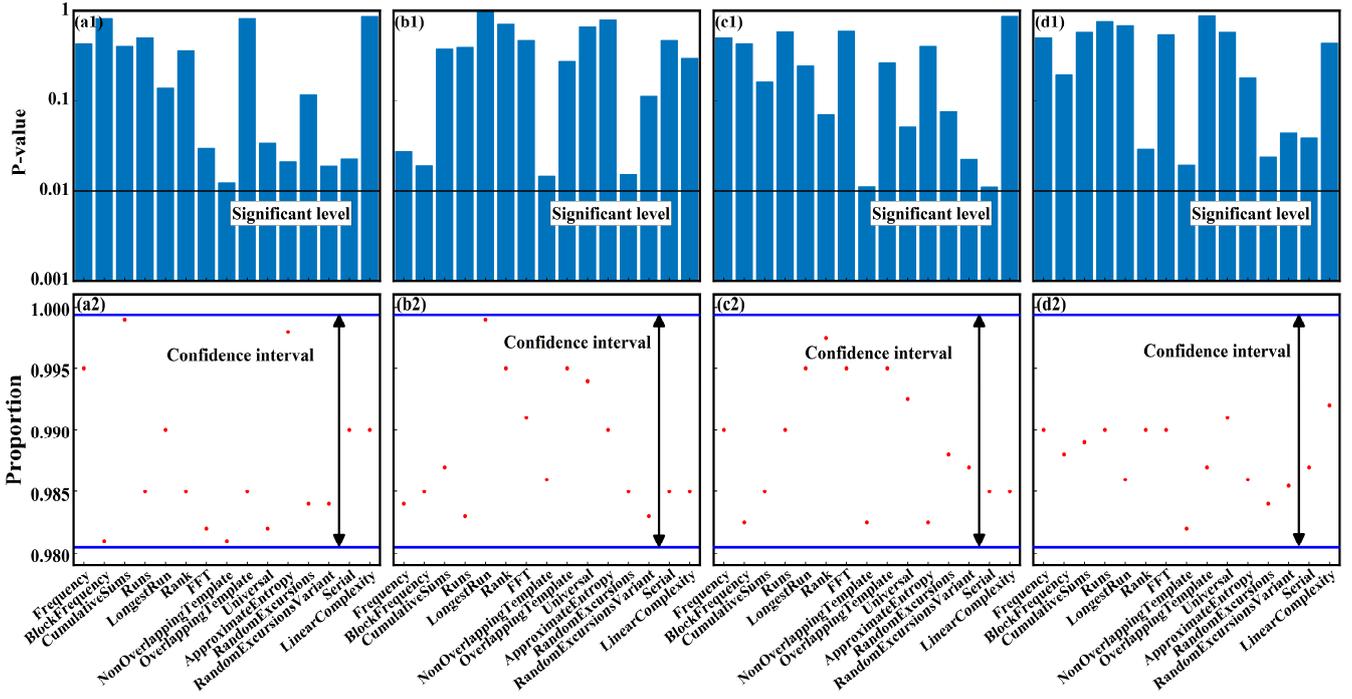

**Figure 6.** (a1) and (a2) ((b1) and (b2), (c1), and (c2)) are the P values and the minimum pass rates for each statistical test of the random numbers souring form the quantum frequency mode center at 200 MHz, 600 MHz, 1 GHz. (d1), (d2) is the P value and the pass rate of the ultimate random numbers.

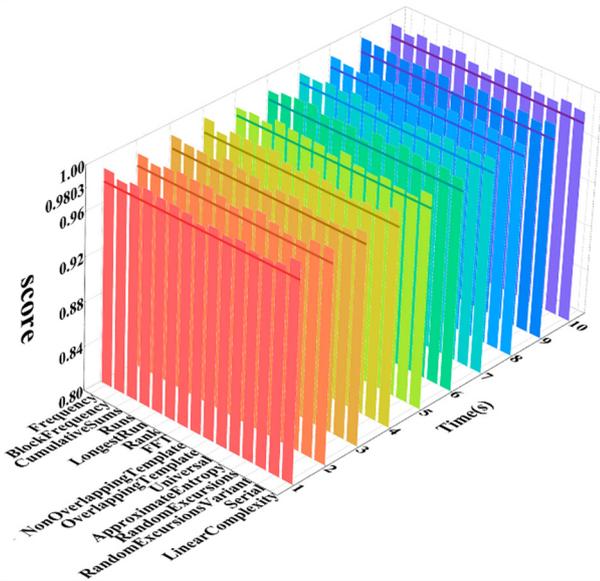

**Figure 7.** The histogram for each color in the graph represents the average minimum pass rate for NIST statistical test of the 8 sets of random numbers collected during the hour.

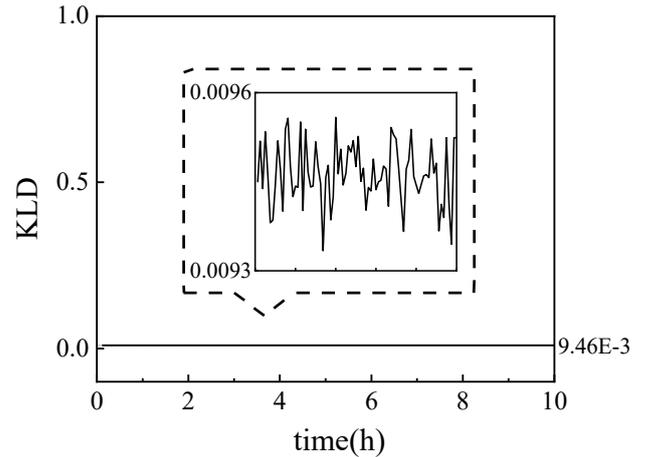

**Figure 8.** The picture shows the timing of the KLD of the data collected within 10 hours. The KLD is close to zero and the fluctuation range is between 0.0093 and 0.0096.

The second test is Kullback-Leibler divergence (KLD) evaluations. The KLD characterizes the proximity of the distribution of two sets of data. The more closely the two sets of data distribute, the KLD is closer to zero. The KLD test takes the first random sequence collected as a reference, and then calculate the KLD of 79 sets of random numbers gathered later and observe the trends. The test results are shown in Fig. 8. The relative entropy almost keeps at a constant level and close to 0 within 10 hours, indicating that the statistical distribution of the generated random numbers remains stable.

## 4. Conclusion

In summary, for realizing multi-channel parallel real-time post-processing of QRNG, we investigate the logic source occupation in a modest FPGA when apply Toeplitz matrices of different scales. A two-layer parallel pipeline algorithm is designed to fully exploit the parallel algorithm advantage and hardware source of FPGA. Three-way Toeplitz post-processing of three sets of raw data from distinct quantum sideband modes is realized with an accumulate rate of 8.24 Gbits/s. The 3-channel post-processing takes up about 61% of the logic resources in the FPGA.

Before and after the post-processing, the autocorrelation and cross-correlation properties are analyzed and the independence between modes are ensured. The generated random numbers are tested by standard NIST test. In addition, we monitor the long-term operation of the system. The Kullback-Leibler divergence of the generated random numbers is stable and the random numbers can past the standard NIST Statistical Test Suite. This work paves an efficient way to overcome the bottleneck in real-time CV QRNG and shows an economic, integrated and extensible design for immediate commercialization.

## Acknowledgements

The authors greatly acknowledge Prof. Tiancai Zhang for scientific discussions and valuable suggestions. The authors also acknowledge the support from the National Natural Science Foundation of China (NSFC) (61875147, 62075154, 61731014), the National Key Research and Development Program of China (2018YFA0306404, 2020YFA0309703), the Key Research and Development Program of Shanxi Province (International Cooperation, 201903D421049), the Scientific and Technological Innovation Programs of Higher Education Institutions in Shanxi (STIP) (201802053, 2019L0131), the Shanxi Scholarship Council of China (SXSCC) (HGKY2019023), and the Program of State Key Laboratory of Quantum Optics and Quantum Optics Devices (KF201905),.